\documentclass[sigconf]{acmart}

\AtBeginDocument{%
  }

\setcopyright{acmlicensed}
\copyrightyear{2026}
\acmYear{2026}
\acmDOI{XXXXXXX.XXXXXXX}
\acmConference[USRW@RecSys '26]{USRW: Workshop on Unified Search and Recommendation at ACM RecSys 2026}{October 2, 2026}{Minneapolis, MN, USA}
\acmISBN{978-1-4503-XXXX-X/2026/10}

\usepackage{booktabs}
\usepackage{multirow}
\usepackage{makecell}
\usepackage{bm}

\begin{document}

\title[Keeping the Index Open]{Keeping the Index Open: The Recommendation-Side Cost of Shared Search and Recommendation}

\author{Theodore Rogers}
\email{theobrog@amazon.com}
\affiliation{%
  \institution{Amazon Web Services}
  \country{United States}
}

\author{Joe Standerfer}
\email{joesta@amazon.com}
\affiliation{%
  \institution{Amazon Web Services}
  \country{United States}
}

\author{Dmitrii Timoshenko}
\email{timodm@amazon.com}
\affiliation{%
  \institution{Amazon Web Services}
  \country{United States}
}

\author{Haoxue Li}
\email{lihaoxue@amazon.com}
\affiliation{%
  \institution{Amazon Web Services}
  \country{United States}
}

\author{Zuhaib Akhtar}
\email{zuhaibak@amazon.com}
\affiliation{%
  \institution{Amazon Web Services}
  \country{United States}
}

\author{Soyoung Yang}
\email{ysoyoung@amazon.com}
\affiliation{%
  \institution{Amazon Web Services}
  \country{United States}
}

\renewcommand{\shortauthors}{Timoshenko et al.}

\begin{abstract}
A shared search-and-recommendation index has to score new items from their features alone, because search has no exploration slot in which to warm them up. On a public log carrying both surfaces over one catalog, $38.6\%$ of query-search impressions in a held-out week show an item never previously shown or visited. On a narrower slice --- engagements where the user, not the catalog, had no prior contact with the item --- the feature-based tower serves that demand at no measurable penalty against $99$ sampled negatives ($0.9595$ Recall@20 against $0.9510$ warm), a parity a lexical baseline shares and a full-catalog check leaves statistically undecided. Dual-encoder (two-tower) retrieval keeps the index \emph{open} in exactly this sense, where an ID-softmax recommender admits a new item only by retraining. We price that openness on the recommendation side --- the half that admits controlled offline evaluation --- against six sequential baselines, each retrained and tuned through five documented rounds on corrected targets (a float32 timestamp bug in a standard library silently reorders the leave-one-out target for $19.7\%$ of users). Warm accuracy costs $5.2\%$ Recall@20 relative ($11.4\%$ NDCG@20) against the strongest retrained baseline in each metric on MovieLens-1M ($3.4$K items); on MIND ($9.4$K items) the gap narrows to $0.8$--$3.6\%$ of the five strongest baselines, though sixth of seven rows. Cold coverage is the payoff, and the shared index's tightest constraint: under a strict cold-start protocol with zero cold-item leakage the content tower reaches $0.172 \pm 0.006$ Recall@20, $1.4\times$ the strongest of the three dedicated methods we could retrain ($0.124 \pm 0.007$) and $3\times$ a training-free floor, with no cold-start-dedicated training stage. The training objective hides a quality--scale trade-off. Swapping a deployed sampled InfoNCE recipe for exact full-softmax cross-entropy (loss, scoring, sampling and schedule move together) lifts test Recall@20 by $54\%$ on MIND-small and $6.9\%$ on MovieLens-1M, one matched pair per benchmark. But the exact objective recomputes the item tower catalog-wide every step and exhausts accelerator memory at a 240K-item catalog in our implementation, the scale that motivates the architecture. Approximate nearest-neighbor search contributes nothing to the gap and retrieval-stage serving cost does not regress against the ID-softmax class; a history-window sweep attributes half the post-recipe remainder. Exact-quality training at catalog scale is the open problem this leaves.
\end{abstract}



\maketitle

\section{Introduction}

Production platforms increasingly serve search and recommendation from one shared item index: one set of pre-computed item embeddings answers both query-initiated retrieval (query tower)~\cite{huang2020embedding} and history-initiated retrieval (user tower)~\cite{covington2016deep, yi2019sampling}. What this shared dual-encoder architecture costs on the \emph{recommendation} side has not been quantified under controlled conditions. We supply that account with retrained baselines under one evaluation convention. The headline finding is a quality--scale trade-off in the training objective: exact full-softmax cross-entropy is the largest accuracy lever we can isolate by paired experiment, yet in our implementation it stops fitting in accelerator memory at the $10^5$-plus-item catalogs a shared index serves (\S\ref{sec:rq2}). We measure the recommendation-side cost of an architecture search-side systems already deploy; search-side training and evaluation of the trained towers is a natural next step. They bind on the search side too, cold coverage most of all. A recommender can \emph{bootstrap} a cold item by exploring (placing it in slates to harvest the interactions that make it warm) and omit it silently meanwhile; search results are held to per-query relevance, so injecting under-served items to collect feedback is not available there. A lexical index covers a just-ingested item from the first query; what a closed-vocabulary retriever cannot do is contribute ranking signal for it, so the dense channel of a hybrid system is dark on new inventory precisely when the lexical channel is least discriminative. Cold items must therefore be scored on content from the first request, and a shared index inherits the stricter requirement (\S\ref{sec:rq3}). The same per-step catalog-linear compute applies to any pipeline that recomputes the shared item tower catalog-wide during training (\S\ref{sec:rq2}), and the serving curve to every consumer of the index (\S\ref{sec:rq4}).

Transformer-based sequential recommenders such as SASRec and BERT4Rec achieve strong accuracy in next-item prediction by applying self-attention over user behavior sequences~\cite{kang2018self, sun2019bert4rec}. These models classify over the item vocabulary (one trained output vector per item ID), a class we call \emph{ID-softmax} models. The production consequence is a \emph{closed index}: one parameter vector per training-time item ID, so unseen items cannot be scored even with rich features at inference~\cite{schein2002methods, volkovs2017dropoutnet} and catalog growth is served only by retraining. The catalog-sized output scan is a second cost, linear in $N$ and dominant at million-item catalogs~\cite{petrov2024pqtopk}, but not an asymmetry: a dual encoder scans its index with the same matmul, so we report serving as a cost ledger (\S\ref{sec:rq4}) and study the closed index as the structural difference.

RetrievalFormer reframes next-item recommendation as retrieval. A transformer user tower encodes the interaction history into a user embedding, a feature-based item tower encodes items from their attributes, and both towers are trained jointly so that the relevance score is a dot product in a shared space,
\[
s(u, i) = f_u(\text{history}_u)^\top f_i(\text{features}_i).
\]
At serving time, top-$K$ recommendations are produced by (approximate) nearest-neighbor search over pre-computed item embeddings~\cite{johnson2019billion, malkov2018efficient}, and new items are scored zero-shot from their features. Neither property is free. We quantify the accuracy the architecture concedes to sequential full-softmax recommenders and why (a training-objective constraint that binds at unified-catalog scale, the $10^5$-plus-item inventories a shared search-and-recommendation index serves), and the capabilities it gains: cold items served through the same feature-encoded tower that search-side matching requires.

Our contributions:
\begin{enumerate}
    \item \textbf{The accuracy cost and its scaling wall.} Paired experiments isolate the sampled-to-full-softmax recipe swap as the largest single lever, and characterize the memory wall that puts the exact objective out of reach at unified-catalog scale in our implementation. Against six baselines retrained on a corrected-target benchmark and tuned through five documented rounds each, RetrievalFormer reaches $94.8\%$ of the strongest on MovieLens-1M ($0.3739 \pm 0.0013$, 3 seeds); a history-window sweep attributes half the remaining gap (\S\ref{sec:rq2}).
    \item \textbf{A strict cold-start stress test: the shared index's binding constraint.} An ID-softmax model's closed vocabulary admits new items only by retraining, so on a shared index the retraining cadence is the item-availability latency for \emph{both} surfaces. Under an item cold-start split with zero cold-item leakage, we retrain and tune three \emph{dedicated cold-start baselines} under the same protocol (DropoutNet, Heater, ALDI). The feature-based tower exceeds the strongest of them on cold Recall@20 by $1.4\times$ and the rest by $1.6$--$1.8\times$, and sits $3\times$ above a training-free content-similarity floor; a fourth dedicated method is cited as prior art only, its implementations having not cleared licensing review (\S\ref{sec:limitations}). A simple frozen score fusion with tuned matrix factorization (3 seeds) sits above all three on both the cold and warm axes (\S\ref{sec:rq3}).
    \item \textbf{Serving cost and protocol diagnostics.} We report exact-scan and IVF-PQ latency with the ANN-vs-exact top-$K$ recall it costs, plus the same measurement on the retrained ID-softmax baseline; the two classes' retrieval-stage cost is symmetric (\S\ref{sec:rq4}). We also report seen-item echo rates for every model --- how many already-seen items appear in the top 20 --- a diagnostic rarely surfaced in model comparisons. Finally, we audit the evaluation targets: float32 timestamp quantization in a standard benchmarking library silently reorders the leave-one-out target (each user's chronologically last interaction) for $19.7\%$ of MovieLens-1M users, so cross-model comparisons on this benchmark are not directly comparable across pipelines, and retraining on the corrected benchmark lowers every baseline score by $0.016$--$0.039$ (\S\ref{sec:setup}, \S\ref{sec:rq1}).
    \item \textbf{The measurement vehicle.} RetrievalFormer: a dual-encoder sequential recommender with an attention-based heterogeneous feature encoder (AttentionFusion) and embedding tables shared across towers, and zero-shot scoring of unseen items, used here for ablation, not as a novelty claim (\S\ref{sec:methodology}).
\end{enumerate}

\begin{figure*}[t]
\centering
\includegraphics[width=0.60\textwidth]{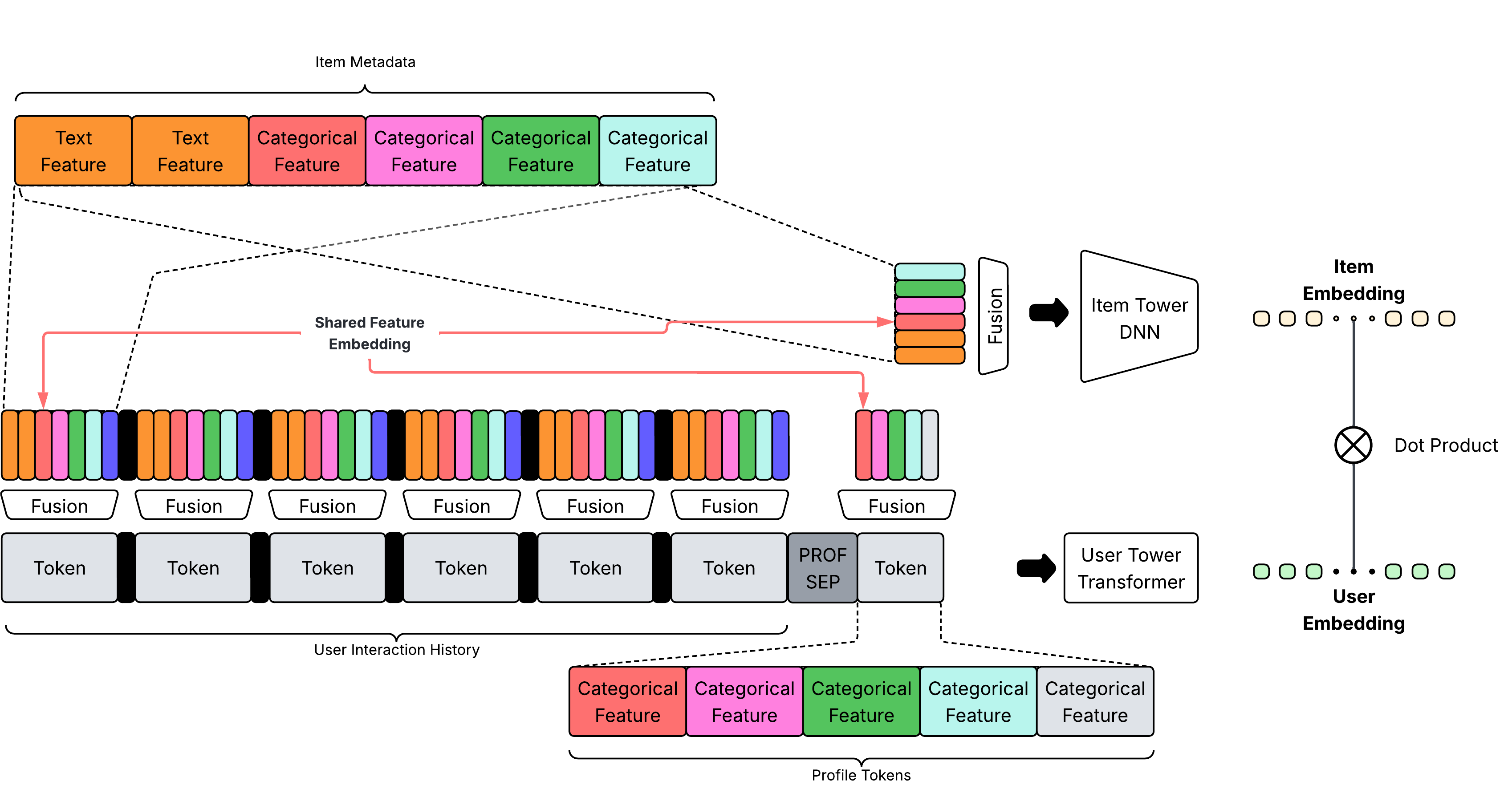}
\Description{Block diagram. Item metadata features are embedded through tables shared with the user side, pooled by a fusion module, and passed through an item-tower network into an item embedding. Each user history event's features are fused into one token; the token sequence, followed by a separator and a fused profile token, feeds a user-tower transformer that produces a user embedding. The two embeddings meet in a dot product.}
\caption{RetrievalFormer data flow. Shared embedding tables (red) embed each categorical value identically wherever it appears: in item metadata, inside a history event, or as a user attribute. Item side: AttentionFusion pools the item's embedded features and the item MLP produces a pre-computable, indexable item embedding ($f_i$). User side: each interaction is fused into one \emph{enriched token}; the sequence [history, [SEP], profile token] feeds the user-tower transformer (prefix-LM: bidirectional within the appended special-token block, causal over history), read out at a final readout token (not drawn) into $f_u$. Relevance is the dot product. The item-ID input pathway with its tied output head (\S\ref{sec:setup}) is not drawn.}
\label{fig:model-architecture}
\end{figure*}

\section{Related Work}

\textbf{Sequential recommendation and transformers.} Sequential recommenders model user-item interaction sequences to predict the next item. Early approaches used Markov chains or RNNs~\cite{hidasi2015session, li2017neural, tang2018personalized, wu2017recurrent}; recent methods are dominated by self-attention. SASRec~\cite{kang2018self} uses unidirectional transformer layers; BERT4Rec~\cite{sun2019bert4rec} uses bidirectional attention with a Cloze objective. These models predict via a softmax over the item vocabulary; the head serves as inner-product scoring, but its vocabulary is closed: items enter it only by training~\cite{su2023beyond}.

\textbf{Two-stage retrieval.} Industry-scale systems retrieve candidates with lightweight models (matrix factorization, two-tower networks) before precise ranking~\cite{covington2016deep, yi2019sampling, huang2020embedding, eksombatchai2018pixie, grbovic2018real}. PinnerFormer~\cite{pancha2022pinnerformer} is the closest deployed relative of our setup, a transformer user tower trained with sampled softmax over pretrained item embeddings for retrieval at Pinterest, reported through deployment metrics, not controlled offline comparison. Sampled and corrected softmax objectives for large-corpus retrieval~\cite{krichene2020sampled, yang2020mixed} are central in this setting. For single-tower ID-softmax models loss choice is decisive: switching SASRec from sampled BCE to full cross-entropy flips reported model rankings on standard benchmarks~\cite{klenitskiy2023gold}, negative sampling induces correctable overconfidence~\cite{petrov2023gsasrec}, and sampled softmax with enough corrected negatives approaches the full objective~\cite{wu2024sampledsoftmax}. A single dot-product similarity head has a separate, documented expressiveness ceiling~\cite{zhai2023revisiting, su2023beyond}. We measure the objective effect in the dual-encoder setting, where the exact objective also hits a catalog-bound feasibility wall (\S\ref{sec:rq2}).

\textbf{Unified search and recommendation.} Search and recommendation are also increasingly modeled jointly: a single model can learn retrieval and recommendation from the same user--item interactions~\cite{zamani2020joint}, and search-behavior signals lift sequential recommendation when the two behavior streams are disentangled or their transitions modeled explicitly~\cite{si2023sesrec, shi2024unisar}. Other work unifies search and recommendation \emph{generatively}: semantic-ID models autoregressively decode item identifiers~\cite{rajput2023generative}, replacing catalog-wide scoring with constrained decoding over compact codeword vocabularies, restructuring exactly the catalog-linear term behind the memory wall of \S\ref{sec:rq2}, and hybrid designs interpolate between dense and generative retrieval~\cite{yang2024unifying}. Our study prices the dual-encoder architecture under controlled retrained-baseline protocols.

\textbf{Attribute-enriched and cold-start recommendation.} Side information improves recommendation under sparsity~\cite{zhou2022filter, de2021transformers4rec, pancha2022pinnerformer}. AttrFormer~\cite{liu2025attrformer} models item attributes inside an ID-softmax transformer; like all ID-softmax models, it cannot score unseen item IDs. UniSRec~\cite{hou2022unisrec} learns transferable sequential representations from item text, sharing the premise that content-based encoders unlock unseen items; we price that premise's dual-encoder serving form under controlled protocols. Dedicated item cold-start methods score cold items from content: DropoutNet~\cite{volkovs2017dropoutnet} trains with representation dropout, Heater~\cite{zhu2020heater} mixes randomized training with expert transformations, and contrastive and distillation variants followed~\cite{wei2021clcrec, huang2023aldi}. Comparing a feature-based model only against ID-softmax models (which trivially score zero on cold items) would overstate its contribution, so we retrain DropoutNet, Heater, and ALDI under our strict split and tune each through documented rounds (\S\ref{sec:rq3}).

\textbf{Evaluation rigor.} Reported gains often shrink against properly trained baselines~\cite{ferraridacrema2019}, so we retrain and tune every baseline under one protocol and report a seen-item echo diagnostic quantifying the evaluation-time masking choice that materially moves all models' scores (\S\ref{sec:setup}).

\textbf{Approximate nearest neighbor search.} IVF, HNSW, and PQ enable millisecond vector search over huge collections~\cite{johnson2019billion, malkov2018efficient}. We propose no new index; we measure what an aggressive IVF-PQ configuration costs in fidelity against exact scoring of the same embeddings (\S\ref{sec:rq4}).

\section{Methodology}
\label{sec:methodology}

RetrievalFormer is a dual-encoder architecture with asymmetric towers (Figure~\ref{fig:model-architecture}). The item tower $f_i(\cdot)$ encodes each item's heterogeneous features into a dense embedding that can be pre-computed and indexed.

 The user tower $f_u(\cdot)$, a transformer, processes the interaction sequence into a query embedding. Recommendations maximize the dot product between the two. Notation: we write $d$ for the shared output embedding dimension in which user and item vectors meet; the transformer width $d_{\text{model}}$ and per-feature embedding sizes $d_f$ are separate hyperparameters (configurations in \S\ref{sec:setup}).

\subsection{Attention Fusion for Heterogeneous Features}
\label{sec:attention-fusion}

Recommendation data mixes single-valued categorical features (e.g., category), multi-valued features (e.g., tags), and text-derived token features. Given features $\mathcal{F} = \{f_1, \dots, f_M\}$ describing an entity, we embed each feature with shared lookup tables and project to a common dimension:
\begin{equation}
\mathbf{H} = [\mathbf{W}_1\mathbf{E}_{f_1}(f_1); \dots; \mathbf{W}_M\mathbf{E}_{f_M}(f_M)] \in \mathbb{R}^{M \times d_{\text{model}}}.
\end{equation}
AttentionFusion is a Set-Transformer-style encoder block~\cite{lee2019set, vaswani2017attention}: multi-head self-attention over the feature set with residual connections and layer normalization, followed by pooling,
\begin{align}
\mathbf{Z} &= \text{LayerNorm}(\mathbf{H} + \text{MultiHeadAttn}(\mathbf{H}, \mathbf{H}, \mathbf{H})), \\
\mathbf{U} &= \text{LayerNorm}(\mathbf{Z} + \text{FFN}(\mathbf{Z})), \qquad
\mathbf{z} = \text{MeanPool}(\mathbf{U}).
\end{align}
The block is permutation-invariant, handles variable-length inputs, and learns how features interact~\cite{song2019autoint}; the same block fuses item metadata, interaction context, and the user profile.

\subsection{Shared Embedding Design}

Embedding tables are shared across towers: when a categorical value (e.g., a category) appears as an item attribute, inside the user's history, or as a user attribute, it uses the same embedding vector. This reduces embedding parameters (roughly $3\times$ in our implementation), keeps feature semantics consistent between towers, and helps cold-start generalization: a new item's features are interpreted by embeddings already trained from every context in which those features occurred. Item representations trained under the recommendation objective remain indexable for query-initiated retrieval from the same unified catalog; whether they reach competitive ad-hoc relevance without a search-trained query tower is untested here (\S\ref{sec:limitations}).

\subsection{Item and User Towers}

\textbf{Item tower.} $\mathbf{y}_i = \text{AttentionFusion}(\mathcal{F}_i) \in \mathbb{R}^{d}$ depends only on item-side features, so all item embeddings are pre-computed offline and indexed; new items receive embeddings from their features without retraining.

\textbf{User tower.} Each historical interaction is encoded as one input token we call an \emph{enriched token}: the item's fused features concatenated with embeddings of that interaction's own context (rating and temporal fields), projected to $d_{\text{model}}$. The transformer processes the sequence
\begin{equation}
\mathbf{S} = [\mathbf{z}_1, \dots, \mathbf{z}_T, \text{[SEP]}, \mathbf{p}_u, \text{tok}_{\text{out}}],
\end{equation}
where $\mathbf{p}_u$ fuses static user attributes and $\text{tok}_{\text{out}}$ is the readout position whose final representation, projected to $\mathbb{R}^{d}$, is the user embedding. Our best configuration uses prefix-LM attention with a last-token readout and additionally feeds item-ID embeddings as an input feature with a tied output head (\S\ref{sec:setup}); the ID pathway helps warm accuracy and is absent for cold items by construction.

\subsection{Training Objectives}
\label{sec:training}

We train the shared embedding space with one of two \emph{recipes}: a bundle of four co-moving choices --- loss, scoring function, negative sampling, and learning-rate schedule --- that production pipelines change together. \S\ref{sec:rq2} measures their difference.

\textbf{Sampled InfoNCE with mixed negatives.} The standard dual-encoder objective~\cite{oord2018representation, yi2019sampling}: for a batch of $B$ user-item pairs,
\begin{equation}
\mathcal{L}_{\text{InfoNCE}} = -\frac{1}{B}\sum_{i=1}^B \log \frac{\exp(\mathbf{x}_i^\top \mathbf{y}_i / \tau)}{\sum_{j \in \mathcal{N}_i} \exp(\mathbf{x}_i^\top \mathbf{y}_j / \tau)},
\end{equation}
where $\mathcal{N}_i$ contains the in-batch items plus $k$ uniformly sampled catalog items (mixed negative sampling~\cite{yang2020mixed}), and $\tau$ is a temperature. InfoNCE implicitly balances alignment of positive pairs against uniformity of the embedding distribution~\cite{wang2020understanding}, which counteracts representation collapse and suits ANN indexing. Its per-step cost is independent of catalog size.

\textbf{Full-softmax cross-entropy.} When the catalog fits in accelerator memory, we instead score \emph{all} items each step with an unnormalized dot product plus a learned per-item bias and train with cross-entropy over the full catalog. This optimizes the full ranking directly. Its per-step memory scales with catalog size, because the entire item tower output (with gradients) is materialized every step; \S\ref{sec:rq2} shows this becomes infeasible surprisingly early.

\section{Experiments}
\label{sec:experiments}

We ask: \textbf{RQ1} How close does RetrievalFormer come to retrained ID-softmax transformers on warm next-item accuracy? \textbf{RQ2} Which measured factors contribute to the remaining gap, and how much remains unattributed? \textbf{RQ3} What does the feature-based item tower deliver under a strict item cold-start protocol, against dedicated cold-start baselines? \textbf{RQ4} Does approximate search contribute to the accuracy gap, what does it cost in fidelity, and how does serving cost compare against the ID-softmax class?

\subsection{Setup and Protocol Diagnostics}
\label{sec:setup}

\textbf{Data.} MovieLens-1M with 5-core filtering: 6{,}040 users, 3{,}416 items, 999{,}611 interactions; leave-one-out splits (last item test, second-to-last validation). For the paired objective experiment in RQ2 we additionally use the MIND news dataset~\cite{wu2020mind} (small edition, ${\sim}$9K interacted items after preprocessing), whose item text (titles, abstracts) exercises the feature-based tower; the paired comparison holds data, features, architecture, and training budget identical within that benchmark.

\textbf{Metrics and convention.} Full-catalog ranking, Recall@20 and NDCG@20, no negative sampling at evaluation on the three recommendation benchmarks (the search-side read of \S\ref{sec:rq3} is the one exception, labelled in place). Echo@20 (the average count of already-seen items in a model's top 20) is reported alongside as a protocol diagnostic. We do \emph{not} mask previously seen items from the candidate set (the common convention for these baselines); Table~\ref{tab:main-results} reports it per model, and \S\ref{sec:rq1} quantifies what the convention choice means for scores. All models are compared under the identical convention; we call this the \emph{exact convention} (exhaustive full-catalog scoring, seen items retained as candidates), and RetrievalFormer's values are exact-convention rescores of its training checkpoints (Table~\ref{tab:main-results} note). Disabling a residual candidate mask in its harness lowers the affected RetrievalFormer scores to the values reported here; it alters no model weights and no baseline score.

\textbf{Target-definition audit.} Leave-one-out's ``last item'' is not framework-invariant: RecBole 1.2.1 downcasts timestamps to float32 on load, which ties same-sitting interactions at MovieLens epochs (${\sim}64$--$128$\,s resolution) and resolves the ties by stable sort over file order, so its \emph{realized} LOO target differs from the chronologically last item for $1{,}190/6{,}040$ ($19.7\%$) of ML-1M users. We verified the mechanism exactly: a float32-quantized stable sort reproduces RecBole's realized targets for every user. Our pipeline's targets are certified chronologically-last (under exact integer timestamps; agreement asserted by build gates) for $100\%$ of users on every benchmark we print, on both test and validation splits. The baselines of Table~\ref{tab:main-results} are retrained end-to-end on a corrected benchmark whose build gate asserts realized target $=$ chronological last for $6{,}040/6{,}040$ users on both splits. Every row of the table therefore shares one target definition. \S\ref{sec:rq1} quantifies what the correction changes: every baseline score falls ($-0.016$ to $-0.039$), and intermediate checkpoint re-scores bound the components.

\textbf{Baselines.} We retrain every baseline ourselves in RecBole~\cite{zhao2021recbole} with full-softmax cross-entropy, then tune each through five documented rounds: a published-config retrain (R0) followed by one configuration lever per round (learning rate, dropout, capacity, masking ratio, batch size), selected on validation NDCG@10 with ties retained by the incumbent, each round recorded in a per-model ledger. Champions on the corrected benchmark: the ID-only sequential models GRU4Rec~\cite{hidasi2015session} (lr $2{\times}10^{-3}$), SASRec~\cite{kang2018self} (lr $5{\times}10^{-4}$, dropout $0.3$), and BERT4Rec~\cite{sun2019bert4rec} (lr $5{\times}10^{-4}$, mask $0.2$), and three side-information sequential models consuming item genres: SASRecF (RecBole's feature-fused SASRec; lr $5{\times}10^{-4}$, dropout $0.3$), FDSA~\cite{zhang2019fdsa} (lr $5{\times}10^{-4}$), and DIF-SR~\cite{xie2022decoupled} at its published $1{\times}10^{-4}$ (the swept base rate collapses it on every benchmark we tested; Table~\ref{tab:main-results} note). Baseline evaluation is validated by an independent NDCG@10 re-scoring self-check (absolute deltas $\le 3.4{\times}10^{-5}$ against the training harness's aggregate). Baseline table rows are 3-seed means (FDSA: single run; Table~\ref{tab:main-results} note); RetrievalFormer's table row is likewise a 3-seed mean; unless stated otherwise, $\pm$ denotes sample standard deviation over the stated seeds. Input parity: no model in Table~\ref{tab:main-results} uses title text; the ID-only baselines consume item IDs alone, SASRecF, FDSA, and DIF-SR add item genres, and RetrievalFormer adds item attributes and user metadata. The information asymmetry between RetrievalFormer and the ID-only baselines favors our model. Title ablations are architecture-dependent, over a complete matrix of three architectures by three benchmarks (nine matched single-seed pairs, $-0.005$ to $+0.010$): titles lift FDSA, the one baseline with a dedicated feature-sequence attention branch, by $+0.0115$ ($0.3767 \to 0.3882$, single seed, replicating in direction on a second MovieLens benchmark), and are inert-to-harmful for SASRecF ($0.3390 \to 0.3372$ on MIND) and DIF-SR (within $\pm0.002$) on every bed. Every title delta is an order of magnitude under the same models' tuning gains ($+0.03$ to $+0.11$ on MIND-large), so token-sequence titles are a second-order feature for these architectures, where the dual encoder reads the same text through a pretrained encoder. Table~\ref{tab:main-results} keeps every baseline at its best no-title configuration to preserve the stated input parity; text content pays off most unambiguously in cold-start (\S\ref{sec:rq3}). AttrFormer~\cite{liu2025attrformer}, an attribute-aware ID-softmax transformer whose implementation is not publicly available, reports $0.413$ Recall@20 on MovieLens-1M; we cite it as reported.

\textbf{RetrievalFormer configuration.} The submitted configuration, selected from sweeps over schedule, dropout, learning rate, batch size, capacity, depth, and history cap, uses prefix-LM attention, last-token readout, and item-ID input with a tied output head. Architecture: $d_{\text{model}}{=}256$, FFN 512; AdamW (weight decay $5{\times}10^{-4}$, none on embeddings) with per-tower learning rates (user $3{\times}10^{-4}$, item $1{\times}10^{-3}$) on a cosine schedule, best-validation checkpoint, bf16. The single-rate constant-LR variant scores $0.3668 \pm 0.004$ (5 seeds) and carries the ablations of \S\ref{sec:rq2}. Trained with the full-softmax objective (\S\ref{sec:training}) and feature-noising regularization\footnote{Feature noising: during training, $5\%$ of observed categorical feature values are replaced with the reserved unknown token before embedding lookup (element-wise for multi-valued bags), so the unknown-token embedding trains regularly and the partially-unknown cold-item case is in-distribution at inference~\cite{srivastava2014dropout, devlin2018bert}.}; the sampled-objective counterpart appears in RQ2.

\subsection{RQ1: Warm Accuracy Against Retrained Baselines}
\label{sec:rq1}

\begin{table}[t]
\caption{MovieLens-1M, full-catalog unmasked ranking, own runs only, all rows sharing one target definition on the corrected chronological-last benchmark (\S\ref{sec:setup}). Echo@20 = average count of already-seen items in the top 20, a protocol diagnostic (\S\ref{sec:rq1}); RetrievalFormer's is scored at seen-window parity against the user's 50 most recent training items, the window the baselines consume. Mean$\pm$std over 3 seeds, each baseline at its best of a published-config retrain plus five validation-selected tuning rounds; $^\dagger$FDSA is a single run (its 3-seed set exists only for the genre+title variant, $0.3838 \pm 0.0043$, which the no-title input-parity rule excludes). RetrievalFormer's seeds are truly-unmasked exact-convention GPU rescores of its shipped checkpoints (per-seed Recall@20 $0.3748$/$0.3745$/$0.3724$). DIF-SR prints its published-learning-rate configuration, which maximizes test Recall@20; its validation-selected round trades Recall for NDCG ($0.3881 \pm 0.0024$/$0.1910 \pm 0.0014$), and ratio statements in the text use the conservative $0.3944$ (against the validation-selected round the warm cost would be $3.7\%$, not $5.2\%$). Best per column \textbf{bold}, runner-up \underline{underlined}.}
\label{tab:main-results}
\centering
\small
\setlength{\tabcolsep}{4pt}
\begin{tabular}{lccc}
\toprule
Model & Recall@20 & NDCG@20 & Echo@20 \\
\midrule
GRU4Rec & 0.3687$\pm$0.0039 & 0.1799 & 3.11 \\
SASRec & \underline{0.3893$\pm$0.0024} & 0.1876 & 3.87 \\
BERT4Rec & 0.3453$\pm$0.0024 & 0.1542 & 4.22 \\
SASRecF (genres) & 0.3851$\pm$0.0015 & \textbf{0.1911} & 3.33 \\
FDSA (genres)$^\dagger$ & 0.3767 & 0.1742 & 4.12 \\
DIF-SR (genres) & \textbf{0.3944$\pm$0.0019} & \underline{0.1883} & 3.99 \\
\midrule
RetrievalFormer (history 300) & 0.3739$\pm$0.0013 & 0.1693 & 4.42 \\
\bottomrule
\end{tabular}
\end{table}

RetrievalFormer recovers most of the strongest baseline's warm accuracy. It reaches Recall@20 $0.3739 \pm 0.0013$ (3 seeds) against retrained-and-tuned baselines (Table~\ref{tab:main-results}): $94.8\%$ of the strongest (DIF-SR, $0.3944 \pm 0.0019$; the NDCG@20 cost is larger, $0.1693$ against SASRecF's $0.1911$, $11.4\%$), within $0.003$ of FDSA ($0.3767$) and $0.011$ of SASRecF ($0.3851$), above GRU4Rec ($0.3687$) and BERT4Rec ($0.3453$), and $0.015$ under tuned SASRec ($0.3893$), the model tuning moved most ($+0.04$ over its published-config retrain). All baselines are retrained and tuned under one documented protocol rather than copied from prior reports: retraining with the full-softmax objective moves them substantially, consistent with the loss-choice findings of~\cite{klenitskiy2023gold}, and the five tuning rounds move four of the six again ($+0.009$--$0.040$). The remaining gap accompanies an architecture whose item side is a feature function, open to unseen items; RQ2 separates its measured recipe and history-context legs from a small unattributed remainder, and RQ3/RQ4 measure what the architecture buys.

\textbf{The correction, quantified.} Retraining end-to-end on the corrected benchmark is the definitive comparison, and it moves the standings: every baseline falls by $0.016$--$0.039$ (published-configuration retrains, R0: GRU4Rec $0.3581$, SASRec $0.3498$, BERT4Rec $0.3296$, SASRecF $0.3757$, FDSA $0.3767$, DIF-SR $0.3924$), the ordering among the baselines shifts (FDSA overtakes SASRecF and GRU4Rec), and the model-vs-baseline ordering changes: our model moves from exceeding one baseline to exceeding three of the six. Those counts are against the published-configuration retrains (R0); the tuning rounds behind Table~\ref{tab:main-results} then lift four of the six corrected baselines back up by $+0.009$--$0.040$, and the standings of record are Table~\ref{tab:main-results}'s. Two diagnostics confirm the direction without being clean reference values, since the affected targets were \emph{training positives} for the original checkpoints: on the original realized targets the same recipe scores higher (GRU4Rec $0.3967$, SASRec $0.3856$, BERT4Rec $0.3460$, SASRecF $0.3997$, FDSA $0.3949$, DIF-SR $0.4177$), under which our model reaches only $89.5\%$ of the strongest; re-scoring those original checkpoints on corrected targets moves them heterogeneously (GRU4Rec $0.3281$, SASRecF $0.3707$, SASRec \emph{rises} to $0.3887$), while a control re-score of our own model measures $+0.0065$ the other way, a small pure-easiness term. The audit procedure is deterministic (\S Reproducibility).
\textbf{Protocol sensitivity.} Echo is universal across the Table~\ref{tab:main-results} families (baseline roster $3.1$--$4.2$ seen items in the top 20; RetrievalFormer $4.42$ at seen-window parity) and is a protocol lever: Masking lifts every model by $+0.030$ to $+0.061$ Recall@20, and masking \emph{both} sides of a pair closes at most about half of its unmasked gap ($17$ gated two-pass rescores, five benchmarks). Protocol choice moves levels, not rankings: no ordering changes at any masking depth we tested, and masking only each user's five most recent items already recovers $35$--$50\%$ of the full-window gain. Mechanism and cross-benchmark normalization are in Appendix~\ref{app:protocol}.

\textbf{Cross-benchmark generalization check.} The warm gap is benchmark-dependent. On a MIND-large 5-core next-item benchmark ($267{,}545$ users; $9{,}353$ items; mean history ${\sim}11$ events, median $8$, so the history-cap lever of RQ2(b) is structurally absent and our arms' cap of $100$ against the roster's $50$ is inert; every model ranks with an exact full-catalog softmax, BERT4Rec's cloze-style and DIF-SR adding its published auxiliaries), RetrievalFormer scores Recall@20 $0.3368 \pm 0.0017$ (3 seeds; top seed $0.3387$) against the same own-run, tuned RecBole roster (Table~\ref{tab:mind-results}): within $0.8$--$3.6\%$ of the five strongest, the tightest band of any benchmark we measure, though sixth of seven rows, while exceeding BERT4Rec ($0.3275 \pm 0.0055$). On AliEC ($30{,}461$ users; $16{,}475$ items after alignment), where our model and the roster both consume a $50$-item history window, it scores $0.0786$ against a leaderboard that tuning kept reordering (top configurations FDSA at $0.0927 \pm 0.0034$ and DIF-SR at $0.0927 \pm 0.0004$, statistically tied; the top spot changed twice across the rounds). MIND-large's tight band is structural: short histories remove the largest single axis of the MovieLens gap, and scoring parity holds on both printed benchmarks (the memory wall of \S\ref{sec:rq2}(d) was measured on the separate raw 240K-item AliEC catalog, for which no head-to-head row is printed).

\begin{table}[t]
\caption{MIND-large 5-core next-item benchmark ($267{,}545$ users; $9{,}353$ items): full-catalog unmasked Recall@20, own runs under the same retrain-and-tune protocol as Table~\ref{tab:main-results}, with the same genre-consuming inputs. Mean$\pm$std over 3 seeds; FDSA is a single run. Short histories (mean ${\sim}11$ events, median $8$) remove the history-context axis of \S\ref{sec:rq2}(b), and every model ranks with an exact full-catalog softmax rather than a sampled one. Best per column \textbf{bold}, runner-up \underline{underlined}.}
\label{tab:mind-results}
\centering
\small
\setlength{\tabcolsep}{6pt}
\begin{tabular}{lcc}
\toprule
Model & Recall@20 & Seeds \\
\midrule
DIF-SR & \textbf{0.3494$\pm$0.0011} & 3 \\
FDSA & \underline{0.3446} & 1 \\
SASRec & 0.3433$\pm$0.0008 & 3 \\
SASRecF & 0.3430$\pm$0.0003 & 3 \\
GRU4Rec & 0.3396$\pm$0.0016 & 3 \\
BERT4Rec & 0.3275$\pm$0.0055 & 3 \\
\midrule
RetrievalFormer & 0.3368$\pm$0.0017 & 3 \\
\bottomrule
\end{tabular}
\end{table}

\subsection{RQ2: Where the Gap Comes From}
\label{sec:rq2}

\textbf{(a) The training recipe is the largest measured contributor.} The recipe swap moves the score more than any other paired lever. On MIND-small, in a single matched pair with architecture, features, data, and budget held identical, switching from sampled InfoNCE (96 mixed negatives per example: 48 in-batch, 24 uniform, 14 popular, 10 tail; cosine scoring, no log-Q correction) to full-softmax cross-entropy (dot-product scoring with a learned item bias) moves test Recall@20 from $0.1830$ to $0.2822$ ($+54\%$ relative). Exact-convention two-pass rescores validate both arms as unmasked-equivalent to within $0.0006$; loss, scoring, sampling, and schedule move together as deployed on this leg. On MovieLens-1M at a matched earlier configuration (4 layers, one pair), the same swap lifts test Recall@20 from $0.3132$ to $0.3349$ ($+6.9\%$ relative; NDCG@20 $+8.9\%$). Both arms sit at a plateau: a converged relaunch early-stopped at the identical checkpoint. The MovieLens-1M anatomy is explicit. The dissection runs from the matched sampled-objective configuration ($0.3132$) to SASRec's legacy realized-target value ($0.3856$), the endpoint that framed this dissection before the target correction of \S\ref{sec:rq1}: a distance of $0.0724$. Of that, the measured recipe contribution recovers $+0.0217$, configuration sweeping adds $+0.0022$, and $0.0485$ remains as the residual of (b). Against the corrected benchmark's strongest baseline (DIF-SR, $0.3944$) the distance is $0.0812$ and the same legs attribute $0.0239$ of it, so the unattributed remainder is bracketed between $0.0117$ and $0.0205$ depending on which endpoint anchors the anatomy; we quote the wider figure where the corrected roster is the comparison. The objective is thus the largest single lever we can isolate by a \emph{paired} experiment; the history-context axis of (b), attributed by sweep rather than pairing, moves more ($+0.0297$ vs.\ $+0.0217$), half the post-recipe remainder against the corrected endpoint; neither is the majority of the gap on this benchmark. Sampled training starves the ordering signal: a given competitor receives gradient with probability ${\approx}0.04$ per query where full-softmax CE updates all $3{,}415$ every step, and NDCG improves proportionally more than recall in both pairs (mechanism, prior art, and two caveats on sampled recipes: Appendix~\ref{app:extended}).

\textbf{(b) A residual survives the axes we swept.} With the full-softmax objective and each swept axis (schedule, dropout, learning rate, batch size, capacity, depth) at its best value at the earlier cap-100 configuration, RetrievalFormer's $0.3371$ trails the legacy-target SASRec by $0.0485$ ($12.6\%$ relative). Measurements bound what this residual is \emph{not}, within the tested bounds: not an evaluation-protocol artifact (masking both sides closes only ${\sim}44\%$ of the checkpoint-pair gap, \S\ref{sec:rq1}), not approximate search (below), not embedding sharing (within seed noise at the optimum; ablation below), and not supervision density. On that last axis, a screen registered before the run replaced the objective with per-position full-softmax cross-entropy on the same architecture and initialization and \emph{lowered} test Recall@20 from $0.3684$ to $0.3260$ (single seed, but $7\times$ the $\pm0.006$ screen threshold). A first direct probe also finds no effect from the item-ID pathway: a single-seed variant trained without the user-side item-ID input or its tied output-side component lands within seed noise of that configuration ($0.3373$ vs.\ $0.3371$). It is \emph{consistent with} the class difference on the item side: the ID-softmax model's item representation is a free per-item parameter vector (pure memorization capacity, updated for every item at every step under full-softmax CE), while ours is constrained to a feature function, which buys zero-shot cold items (\S\ref{sec:rq3}). Dual-encoder capacity analyses in dense text retrieval reach similar conclusions~\cite{luan2021sparse}, and richer accelerator-friendly similarity heads recover expressiveness within the retrieval envelope~\cite{zhai2023revisiting, su2023beyond}. The history-context axis attributes most of it: the earlier pinned configuration caps history at 100 items, covering the median user (median history 96) but discarding $55\%$ of training interactions; raising the cap to 300 (recovering $78.5\%$) lifts the same architecture and recipe to $0.3668 \pm 0.004$ Recall@20 (5 seeds; per-seed range $0.3624$--$0.3725$), with cap 200 ($0.3589 \pm 0.002$, 2 seeds) tracking the coverage curve and cap 400 saturating ($0.3675$, single seed). How much of that window the model needs at \emph{inference} is separable, and smaller: re-scoring the submitted checkpoints with evaluation history truncated to the baselines' $50$ items (padded width held fixed so retained items keep their trained positions) costs $0.3739 \to 0.3671 \pm 0.0023$ (3 seeds, paired, every seed negative). At that context parity the warm gap widens from $-0.0205$ to $-0.0273$ against DIF-SR ($93.1\%$ of its bar, against $94.8\%$ printed), and our ordering against FDSA resolves against us (Welch $t{=}3.4$) where at full window it does not ($t{=}1.1$); against GRU4Rec the two remain indistinguishable at either window ($t{=}0.6$). This caps our model against baselines at their native cap, the fair direction rather than a like-for-like recompute. Roughly three-quarters of the cap's value is therefore recovered training interactions, not inference context. That mechanism is MovieLens-specific: we truncate at ingest, so an interaction outside the window enters no training row, and at each bed's own cap this discards $21.5\%$ of MovieLens-1M interactions against $0.1\%$ on MIND (mean history $10.7$) and $0.2\%$ on AliEC (mean $8.3$), where the window is inert and the asymmetry with the roster's $50$ costs nothing. Splitting the tower learning rates on that configuration (user $3{\times}10^{-4}$, item $1{\times}10^{-3}$, cosine schedule) adds $+0.007$, reaching the submitted $0.3739 \pm 0.0013$ (3 seeds, GPU-verified exact rescores) of Table~\ref{tab:main-results}. On the corrected benchmark (Table~\ref{tab:main-results}) the submitted model sits above the ID-only GRU4Rec and BERT4Rec and $0.003$--$0.021$ under the four strongest (DIF-SR the farthest); measured against the legacy realized-target SASRec value that framed this dissection, the unattributed remainder is $0.012$; loss variants and supervision density remain unexhausted. A cutoff decomposition localizes it: against the strongest baseline (DIF-SR) the ratio is monotone in the cutoff: $86\%$ at Recall@5, $92\%$ at Recall@10, $95\%$ at Recall@20, and seen-item exclusion on both sides closes roughly half of that head gap ($86\%{\to}92\%$ at Recall@5), so the head-of-list deficit is part evaluation echo and part the item-side memorization account above (Appendix~\ref{app:protocol}). Isolating the remainder (further ID-pathway factorization, distillation from candidate-conditioned teachers) is future work.

\textbf{(c) ANN approximation error contributes nothing to the accuracy gap.} Every accuracy number here comes from exact, exhaustive scoring of its full candidate set, so the informal attribution of dual-encoder gaps to ``the ANN approximation'' is ruled out by construction, and a real-embedding check agrees: HNSW on the MIND-large checkpoint reproduces the exact end metric within noise (\S\ref{sec:rq4}).

\textbf{(d) The recoverable leg hits a memory wall at scale.} The exact objective stops fitting in memory before the catalog gets large. The full-softmax objective materializes the entire item tower forward (with gradients) every training step, so its memory footprint carries a term linear in the \emph{catalog} that batch reduction cannot remove: in our implementation on a single 44\,GB GPU it trains at a 3.4K-item catalog (MovieLens-1M) and a 9K-item catalog (MIND) but goes out of memory at a 240K-item catalog at both tested batch sizes (512 and 256); the failing 470\,MiB allocation matches one catalog-wide $d{=}512$ fp32 activation. The constraint is not a kernel artifact. For a plain ID-parameterized output layer, fused cross-entropy kernels eliminate the logit memory outright~\cite{wijmans2025cce}; but a \emph{feature-based} item tower must recompute the catalog-wide tower forward and backpropagate through it every step, so even memory-optimal exact training pays per-step compute linear in catalog size (we did not test gradient checkpointing, activation offload, or sharding, which relax memory but not this compute term), and the practical escapes buy scale by leaving the exact objective behind: caches score against stale embeddings~\cite{lindgren2021negativecache, wang2021cbns} and adaptive-softmax factorizations restructure the loss~\cite{grave2017efficient}. Prior work reaches the same verdict from the ID-softmax side (full softmax impractical at million-item catalogs~\cite{petrov2023gsasrec}); concurrent work engineers around precisely this wall with approximate cross-entropy losses~\cite{mezentsev2024sce, gusak2024rece}. At a 2.2M-item catalog a full $d{=}512$ table is $4.2$\,GiB of parameters and $16.8$\,GiB with gradients and Adam moments, which is why million-item work in this line represents items by \emph{compressed} sub-item embeddings rather than a full per-item table~\cite{petrov2025recjpqprune}. Both classes pay a catalog-linear training cost (theirs in item parameters, ours in the tower forward), so the wall is a property of exact catalog-scale objectives. The exact objective that anchors the recoverable part of the gap is trainable where the dual encoder is least needed (small catalogs) and infeasible at the scales that motivate it. Closing the distance between corrected sampled objectives~\cite{klenitskiy2023gold, wu2024sampledsoftmax, krichene2020sampled} and the full-softmax anchor at catalog scale is an open problem for retrieval-based sequential recommendation.

\textbf{Within-class ablations (Appendix~\ref{app:extended}).} Sharing embedding tables across towers matters on average but not at the optimum (marginalized over the tuning grid: separate tables cost $3.4\%$ Recall@20, no shared features $6.9\%$; within seed noise at the best configuration), robustness, not an accuracy driver. Replacing prefix-LM attention with pure causal attention costs $-0.0098$/$-0.0026$ on paired seeds, so the bidirectional prefix block is load-bearing at the margin and the submitted configuration keeps it.

\subsection{RQ3: Strict Item Cold-Start}
\label{sec:rq3}

\textbf{Protocol.} We use a strict item cold-start split with zero leakage: 20\% of items (742 of 3{,}706; grouped by item, seed-fixed) are held out as cold, and \emph{every} interaction involving a cold item is removed from training. This benchmark retains all rated MovieLens-1M items; the warm-accuracy benchmark of \S\ref{sec:rq1} is 5-core filtered (3{,}416). Cold items are split into disjoint validation and test sets (371 each); warm interactions are split 8:1:1. Following the benchmark's slice convention, the cold slice ranks all 742 cold items, the warm slice ranks all 2{,}964 warm items, and the overall slice ranks the full catalog, always excluding the user's training items; cold items are candidates for every model, so the cold-start tax is explicit. A global random item holdout samples the catalog uniformly, unlike per-user leave-last-out; exhaustive partition ranking, unlike sampled-negative AUC, is comparable to the strict cold-start literature~\cite{volkovs2017dropoutnet, zhu2020heater, wei2021clcrec, huang2023aldi}. Appendix~\ref{app:coldstart-protocol} details splits and baseline training. Each baseline is retrained to convergence at 3 seeds and tuned through the same five-round, one-lever protocol as the warm baselines (\S\ref{sec:setup}). Analytic floors: a random ranker's cold Recall@20 is $20/742 \approx 0.027$, and a popularity ranker is uninformative on cold items (all have zero training interactions). We add a behavioral floor: ranking candidates by cosine similarity between the frozen text embedding of the user's last training item and each candidate's embedding (zero training, byte-identical frozen features to our item tower), reaches $0.058$ cold Recall@20, a floor any \emph{trained} cold-start method should clear.

\begin{table}[t]
\caption{Strict item cold-start on MovieLens-1M (global random 20\% item holdout; cold/warm slices rank their candidate partitions, overall ranks the full catalog). Mean$\pm$std over 3 seeds where shown; row definitions and protocol in \S\ref{sec:rq3} and Appendix~\ref{app:coldstart-protocol}. MF and the dedicated baselines are tuned through five documented one-lever rounds with 3-seed champion confirms. Item content is frozen sentence embeddings (``text'') or genre+year (``meta''); text variants remain at reference configurations. Best per column by point estimate \textbf{bold}, runner-up \underline{underlined}; ``---'' = not computed. The three bolded warm cells are the same tuned-MF ceiling, which the fused rows inherit by construction.}
\label{tab:coldstart}
\centering
\footnotesize
\setlength{\tabcolsep}{3pt}
\begin{tabular}{lccc}
\toprule
Model & Cold R@20 & Warm R@20 & Overall \\
\midrule
\multicolumn{4}{@{}l}{\textit{Floors}} \\
\quad Last-item text cosine (no training) & 0.058 & 0.018 & --- \\
\quad MF (ID-based CF, tuned) & 0.031$\pm$0.004 & \textbf{0.276$\pm$0.001} & \textbf{0.125} \\
\midrule
\multicolumn{4}{@{}l}{\textit{Dedicated cold-start baselines (tuned)}} \\
\quad DropoutNet (meta) & 0.094$\pm$0.013 & 0.241$\pm$0.001 & 0.109 \\
\quad Heater (meta) & 0.106$\pm$0.005 & 0.256$\pm$0.002 & 0.116 \\
\quad ALDI & \underline{0.124$\pm$0.007} & \underline{0.267$\pm$0.000} & \underline{0.121} \\
\midrule
\multicolumn{4}{@{}l}{\textit{RetrievalFormer: varying how a warm channel is attached}} \\
\quad Content tower, no warm channel & \underline{0.172$\pm$0.006}$^\ddagger$ & 0.113$\pm$0.005 & 0.067 \\
\quad \,$+$ warm-masked CE (retrained) & 0.154$\pm$0.003 & 0.131$\pm$0.005 & --- \\
\quad $\oplus$ tuned MF (score fusion) & 0.172$\pm$0.006$^\ddagger$ & \textbf{0.276$\pm$0.001} & \textbf{0.125}$\pm$0.000 \\
\quad $\oplus$ tuned MF, instr.\ encoder & \textbf{0.195$\pm$0.008} & \textbf{0.276$\pm$0.001} & \textbf{0.125}$\pm$0.000 \\
\bottomrule
\end{tabular}

\vspace{2pt}
{\footnotesize $^\ddagger$These two cold values are equal by construction: MF's cold rows are zeroed and the fusion scalar is cold-slice invariant, so the same frozen content tower is the sole cold scorer in both rows. Only the warm channel differs.}
\end{table}

\textbf{Results (Table~\ref{tab:coldstart}).} The ID-based CF model collapses on cold items ($0.031 \pm 0.004$, within noise of the $0.027$ chance floor). The dedicated cold-start baselines clear the CF floor, and tuning moves them substantially (Heater $0.087 \to 0.106$, DropoutNet $0.077 \to 0.094$; ALDI reaches $0.124$), but DropoutNet's and Heater's \emph{text} variants land at the training-free last-item floor ($0.054$--$0.059$ vs.\ $0.058$; Appendix~\ref{app:extended}): strong content alone does not suffice. RetrievalFormer's content-only tower reaches $0.172 \pm 0.006$ cold (3 seeds): $1.4\times$ ALDI, $1.6\times$ tuned Heater, $3\times$ the floor, $5.5\times$ tuned MF, from the retrieval architecture without a dedicated cold-start training stage. These margins are over the three methods we could retrain; we omit CLCRec (\S\ref{sec:limitations}). The warm-masked full-softmax variant (the same recipe with cold items masked out of the training softmax, so the loss trains on warm targets only; 3 seeds) trades cold for warm ($0.154 \pm 0.003$ / $0.131 \pm 0.005$), consistent with RQ2(a): the recipe arc from sampled InfoNCE lifts warm from ${\sim}0.07$ to $0.131$ at little cold cost, closing under a third of the deficit to the tuned-MF ceiling, and raising the learning rate toward its warm optimum monotonically degrades cold retrieval. An encoder swap is not automatically the lever: substituting a newer, MTEB-stronger 0.6B text-embedding model into the champion recipe unchanged \emph{lowers} cold recall by $19\%$ relative --- a statement about reusing a recipe tuned for one embedding geometry, not about encoder quality, since we did not re-tune under the new encoder.

\textbf{A fused cold/warm operating point.} A frozen score fusion reaches both axes at once. Fusing the frozen content encoder with the frozen \emph{tuned} MF of Table~\ref{tab:coldstart} (zeroing MF's cold rows; one scalar weight per content seed, selected on warm-slice validation; the cold slice is $s$-invariant by construction) attains cold $0.172 \pm 0.006$ / warm $0.276 \pm 0.001$ (3 seeds): by construction cold ranking is carried entirely by the content component and warm saturates at the backbone's level, so the fusion carries the content tower's cold at the tuned-MF warm ceiling, above every dedicated baseline's 3-seed row on both axes. The overall slice is $80\%$ warm items, so it tracks warm capacity: the fusion lifts the content tower from $0.067$ to $0.125$ there, the best overall value in the table. Attaching a frozen CF backbone to a content encoder is the construction the dedicated baselines use (DropoutNet, Heater, and ALDI are all built on one); the fusion differs in combining at score level rather than in a learned layer, and in requiring no additional training. The fusion holds across content models. With an instruction-tuned encoder variant, the same protocol reaches cold $0.195 \pm 0.008$ at the same warm ceiling ($0.276 \pm 0.001$; 3 seeds, one test read at the validation-selected weight), the strongest fused point we measure. Two caveats: the win-both operating point is the fusion (the DropoutNet/Heater recipe with a stronger content encoder and a properly trained CF backbone), and \emph{co-training} a single model to win both did not, across five single-seed hybrid arms, though objective isolation does not require separate serving models. We train the memorization (ID-embedding CF) channel in isolation and select its endpoint on validation: a 3-point training-budget curve whose later endpoints overfit the memorization table, ranked correctly by validation, so the choice involves no test peeking. We then \emph{graft} the content champion's weights under it; the two parameter groups are disjoint, blended by the same validation-selected scalar protocol as the fusion. The result is a single valid checkpoint at cold $0.172 \pm 0.006$ / warm $0.250 \pm 0.000$ (3 content seeds under the same channel): cold fully held, warm at $90\%$ of the tuned-MF ceiling, zero additional training. A sanity gate confirms the grafted content path reproduces the champion's scores per seed. Co-training did not win both; training the channels separately and composing them does. This composition carries two-run provenance.

\textbf{Two controls (Appendix~\ref{app:extended}).} Giving DropoutNet and Heater the same frozen text embeddings \emph{drops} their cold recall (Heater $0.087 \to 0.054$), so the advantage tracks how the architecture uses the encoder, not the encoder itself; and the warm deficit is protocol-relative, each class winning its matched task ($0.267$ vs.\ our $0.113$ on matrix completion, $0.350$ vs.\ MF's $0.125$ on leave-one-out). Only the feature-based tower also wins cold.

\textbf{Cold coverage on the search surface.} The openness premise is a search-side claim, so we
test it there, on the Avito context-ad log: query-initiated search ($112.2$M searches, $15.6\%$
with free-text queries) and organic browsing ($286.8$M events) over one $36.9$M-ad catalog.
Freshness pressure is heavy independent of any label choice --- of $9.77$M query-search
impressions in a held-out week, $38.6\%$ show an ad absent from every earlier impression
\emph{and} visit ($41.0\%$ of distinct ads shown), the premise's antecedent measured rather than
assumed. Organic inventory carries no click labels, so relevance comes from delayed cross-stream
engagement: a shown ad the same user visits $30$ minutes to $24$ hours later ($246{,}443$ pairs;
the floor discards the $81.9\%$ that are SERP click-throughs, median gap $28$\,s, and a placebo
pairing each search with another user's impressions converts $1{,}700\times$ less often).
Against $99$ sampled negatives, on engagements where the user had no prior contact with the item,
so only content can retrieve it, Recall@20 is $0.9595 \pm 0.0008$ cold against
$0.9510 \pm 0.0008$ warm (3 seeds, $12{,}946$/$15{,}368$ instances; query tower alone
$0.9510$/$0.9433$). Two receipts bound that reading. Under the same sampled protocol, untuned BM25 over unstemmed
titles also ranks cold above warm ($0.7596$/$0.7261$ pooled, against the tower's
$0.9685$/$0.9646$), so the slice favors cold items independently of the tower; and on the
undivided test set the strongest model-free scorer we built reaches $0.8791$ against the tower's
$0.9664$, so the sampled task is easy but not saturated. And against the entire $22.8$M-ad pool (single seed; $549$/$638$ of a
$2{,}000$-instance sample fall in this stratum) cold and warm are indistinguishable at every depth
($-0.71$pp Recall@20 $[-1.83, +0.31]$; $+1.86$pp Recall@1000 $[-1.74, +5.46]$). The claim is the absence of a
\emph{measurable} cold penalty, not a cold advantage, and not parity at catalog scale.

\subsection{RQ4: Serving as a Cost Ledger}
\label{sec:rq4}

Exact full-catalog scoring of dual-encoder embeddings holds 90th-percentile (p90) latency under 30\,ms up to 1M items, so approximation is unnecessary at this scale. Within this envelope (1{,}024-query batches over synthetic $d{=}512$ vectors, wider than our $d_{\text{model}}{=}256$, so the scan term is conservative), every accuracy benchmark here is scored exactly, without approximation. Beyond that, exact scan reaches $292\,$ms at 10M items against an aggressive IVF-PQ index's ${\sim}1\,$ms, which retrieves only $2.8\%$ of the exact top-$K$ ($m{=}16$, $n_{\text{probe}}{=}16$). A real-embedding check closes the loop between index recall and the end metric: on the certified MIND-large checkpoint ($9{,}353$ items, $267{,}545$ test users; the learned item bias folded in by coordinate augmentation, so index inner product equals the exact score), HNSW ($M{=}32$, depth 100) reproduces exact-scan Recall@20 within noise ($0.33878$ vs.\ $0.33884$, $99.98\%$ retention) and IVF-Flat retains $99.86\%$, but at this scale ANN is not a latency win: exact CPU scan is already $0.28$\,ms p90 per query, and the user-tower forward (${\sim}5$\,ms) dominates regardless of retriever. Approximation becomes load-bearing only above ${\sim}10^6$ items, where the batched scan reaches tens of milliseconds per query; fidelity can then be bought back (raising $n_{\text{probe}}$ from $8$ to $16$ lifts recall from $0.021$ to $0.028$ for $+0.1$\,ms at 10M items). Serving cost is symmetric across the classes, and scale-dependent. Timing the retrained SASRec champion's own scoring path (single-thread CPU, $d{=}256$) puts the sequence forward at $94$--$99.5\%$ of its \emph{full\_sort\_predict} time (encoder $2.2$--$3.0$\,ms over that call's $2.3$--$3.0$\,ms, batch sizes 1 and 1024), with the scan timed in isolation at $0.02$--$0.09$\,ms, at this benchmark's $3{,}416$ items plus the padding row; the same scan on synthetic catalogs inverts that ratio, the encoder falling to $27\%$ of cost at $10^6$ items and $4\%$ at $10^7$. Both classes run one sequence forward and one scan against a static table, so the profile is theirs equally: at benchmark scale neither is scan-bound, at catalog scale both are. End-to-end ANN impact at $10^6$-plus scale is future work.

\section{Limitations}
\label{sec:limitations}
Warm-accuracy and cold-start grids are on MovieLens-1M; MIND and AliEC enter as paired-objective and cross-benchmark checks (mixed single-run and 3-seed rows, Table~\ref{tab:mind-results}). The cross-protocol MF cell is single-seed (the fusion, grafted composition, and warm-masked CE rows are 3-seed; the graft carries two-run provenance); baselines in Table~\ref{tab:main-results} are 3-seed means of each champion configuration, and the cold-start baselines are 3-seed and tuned through the same round protocol (text variants remain at reference configurations). We did not enforce parameter parity, and our model's configuration search was broader than the five rounds each baseline received, a budget asymmetry favoring our model. The baselines also run their framework's default $50$-item history window, and history length was not among their five tuning levers, so the history-context leg of \S\ref{sec:rq2}(b) is an axis our model was swept on and they were not. We did not retrain a MovieLens baseline at a longer window, so on that benchmark we cannot separate that leg from what a baseline would gain from the same window --- only what our own model loses without it (\S\ref{sec:rq2}); on AliEC, where both sides run the $50$-item window natively, we trail the two strongest by $15\%$. The dedicated cold-start roster is three methods rather than four: we omit CLCRec, whose available implementations we could not clear under licensing review, and cite it as prior art only. The residual in RQ2(b) is bounded by elimination and largely attributed by the history-context axis (the submitted configuration is the 3-seed split-LR cap-300 family; the dissection used the 5-seed constant-LR cap-300 family), and the remaining isolations (supervision density, loss variants) have not run. Scan timings beyond $10^4$ items use synthetic embeddings on one host and vary by ${\sim}15\%$ run to run, and absolute latencies are hardware-dependent. The batched benchmark measures the catalog-scan term only, and the quoted IVF-PQ point is an aggressive-quantization, low-recall configuration. The ANN sweep covers one index family at two probe settings and the real-embedding end-metric check one benchmark (MIND-large); end-metric ANN impact at $10^6$-plus-item scale remains unmeasured. The objective effect of \S\ref{sec:rq2}(a) is a recipe bundle, and quality and trainability were not measured across catalog scales on one benchmark. No head-to-head accuracy comparison runs at the $10^5$-plus-item scale that motivates the architecture: the accuracy grids are $3.4$K and $9.4$K items, and the memory wall is measured at 240K without an accompanying accuracy read. A feature-based tower also inherits metadata-quality bias: items with sparse or malformed descriptions receive lower-quality embeddings, a coverage concern for long-tail segments that our method does not correct. The search-side result pairs a label-free prevalence count with a cold-coverage read whose relevance signal is delayed engagement on one external log, and whose lexical comparator is an untuned BM25 rather than a tuned search baseline. Its full-catalog leg is a single model seed, and the $2{,}000$-instance sample leaves $1{,}187$ in the no-prior-contact stratum, so its Recall@20 interval spans about $2$pp: it establishes that no cold penalty is measurable at that power, not that none exists. Training the second tower as a query encoder and measuring blended retrieval against a tuned baseline is future work.

\section{Conclusion}

The exact objective is trainable where the architecture is least needed, and infeasible (in our implementation) at the catalog scales a shared index demands. That trade-off is the central finding. The cost is $94.8\%$ of the strongest retrained baseline, on targets a float32 artifact had been silently reordering (\S\ref{sec:rq1}); in exchange the index stays open, at $1.4\times$ the strongest of the three dedicated methods we retrained, with no retrieval-stage cost regression (\S\ref{sec:rq3}). The memory wall remains open.

\bibliographystyle{ACM-Reference-Format}
\bibliography{references}

\appendix

\section*{Reproducibility Statement}
All datasets are public (MovieLens-1M, MIND, AliEC). Item text is encoded by a frozen public sentence encoder (gte-Qwen2-1.5B, $1{,}536$-dim, L2-normalized); the cold-start encoder-substitution control of \S\ref{sec:rq3} swaps it for a newer $0.6$B model at higher MTEB rank. Baselines are retrained in RecBole with full-softmax cross-entropy, tuned through five documented one-lever rounds each, and evaluated full-catalog unmasked. Cross-harness data and candidate parity was verified directly (identical 5-core statistics and train-interaction counts; a popularity ranker scores 0.0402 unmasked Recall@20 in both stacks), and build gates assert target identity for all $6{,}040$ users on both corrected splits. Large language models were used for literature search and writing assistance only; every reported number originates from a logged training or evaluation run.

\section{Cold-Start Protocol Details}
\label{app:coldstart-protocol}
Interactions are grouped by item and items are shuffled with a fixed seed; 80\% of items (with all their interactions) form the warm pool, split 8:1:1 into train/validation/test by interaction. The remaining 20\% of items contribute \emph{all} of their interactions to the cold pool, split 1:1 by item into disjoint cold-validation and cold-test sets, so no cold item's ID or interactions appear in training and no item is shared between cold validation and cold test. This benchmark is immune by construction to the timestamp-quantization artifact of \S\ref{sec:rq1}: its targets are item-random rather than time-derived (the split procedure contains no timestamp logic), and our pipeline preserves integer timestamps end-to-end (verified over all 646{,}073 events). At evaluation, each slice ranks its candidate partition minus the user's training history: the cold slice all 742 cold items, the warm slice all 2{,}964 warm items, and the overall slice the full catalog, with cold items scored by every model; ID-based models score cold items through whatever mechanism they possess (for MF, the untrained embedding). Cold-start baselines (DropoutNet, Heater, ALDI) are trained on the warm pool with item content (genre+year metadata, or frozen sentence embeddings of titles in the ``text'' variants) and evaluated identically. All three consume the same frozen matrix-factorization backbone. Each ran a published-config retrain plus five documented one-lever tuning rounds (validation-selected; champions confirmed at 3 seeds), under a 500-epoch cap with early stopping; text variants remain at reference configurations (e.g., Heater's 200-unit expert MLP), without per-variant capacity re-tuning. The last-item floor scores every candidate by cosine similarity to the frozen text embedding of the user's most recent training item, using the same frozen embeddings our item tower consumes. Model selection for all cold models uses cold-validation; the fusion scalar $s$ in Table~\ref{tab:coldstart} is selected to maximize warm-slice validation accuracy (under a cold floor that never binds) and applied once to test; the cold slice is invariant to $s$ by construction, since MF's cold rows are zeroed and the cold candidate partition contains only cold items.

\section{Evaluation-Protocol Sensitivity}
\label{app:protocol}

Echo mechanism: on single-consumption benchmarks like MovieLens a seen item can never be the held-out target, so an evaluation-time seen-item filter converts every echoed slot into free rank improvement, hence the paper's unmasked numbers plus the echo diagnostic. The RetrievalFormer echo of \S\ref{sec:rq1} is counted at seen-window parity on the printed history-300 seeds and sits above the top of the baseline band. Nor is echo a MovieLens quirk: on the sparser MIND and AliEC benchmarks, echo runs $15$--$150\times$ the random-ranking expectation ($20 \times$ mean history length / catalog size), and masking flips no model comparison. The ${\sim}44\%$ gap closure quoted in \S\ref{sec:rq2}(b) is the SASRec pair under both-sides masking. Echo is computed from input sequences alone and is independent of the target-definition artifact of \S\ref{sec:rq1}.

Cutoff decomposition (\S\ref{sec:rq2}(b)): the submitted model's NDCG ratios against DIF-SR are $85\%$/$89\%$/$91\%$ at cutoffs $5$/$10$/$20$, and both-sides seen-item exclusion lifts the NDCG@5 ratio $85\%{\to}90\%$. Our side is 3 seeds; the baseline side is single-seed re-scores of the champion checkpoint, so the @20 endpoints differ from the Table~\ref{tab:main-results} mean ratios by about one point.

\section{Extended Analyses}
\label{app:extended}

\textbf{Why sampled training starves the ordering signal (from \S\ref{sec:rq2}(a)).} With the MovieLens-1M arm's $128$ negatives per query against its $3{,}416$-item catalog, a given competitor receives gradient with probability ${\approx}0.04$ per query (under one of a user's current top-20 competitors updated per step), while full-softmax CE updates all $3{,}415$ every step. Its empirical fingerprint is present: NDCG improves proportionally more than recall in both pairs, and the same pathology is well documented for ID-softmax models~\cite{petrov2023gsasrec, klenitskiy2023gold}. Two caveats. First, these are matched-configuration comparisons; we do not claim no sampled recipe can do better; no single loss dominates across datasets and evaluation setups, as measured on sequential ID-softmax models~\cite{logqcorrection2025}. Second, sampled cross-entropy with large corrected negative pools (thousands of negatives) approaches full-softmax quality for ID-softmax models~\cite{klenitskiy2023gold, wu2024sampledsoftmax}, while our arms use the mixed in-batch recipe standard for dual encoders (\S\ref{sec:training}).

\textbf{Encoder-axis control (cold-start).} Giving DropoutNet and Heater the \emph{same} frozen text embeddings RetrievalFormer uses does not close their gap; it lowers their cold recall (Heater $0.087 \to 0.054$, to the level of the training-free floor; DropoutNet $0.077 \to 0.059$) while nudging warm up (DropoutNet text $0.211 \pm 0.002$ warm, Heater text $0.251 \pm 0.003$ warm; 3 seeds). Both baselines run at their reference configurations, whose capacities were designed for low-dimensional metadata rather than $1{,}536$-dim text embeddings, so this control bounds the reference recipes, not the architectures' ceiling. Within that bound the conclusion is consistent across three probes (baselines-with-our-features, the last-item floor on byte-identical features, and our tower): the advantage comes from how the dual-encoder architecture uses the content encoder.

\textbf{What ``losing warm'' means is protocol-relative.} The cold benchmark's warm slice is a matrix-completion task (rank a user's held-out warm ratings) which rewards per-pair memorization, the home game of MF. Running both model classes through one shared harness in both directions illustrates the task-dependence descriptively (the two protocols differ in split, candidate set, and checkpoints; this is not a controlled decomposition): on the cold benchmark's warm slice, MF scores $0.267$ and our content tower $0.113$; on leave-one-out \emph{next-item} prediction, our model scores $0.350$ while MF scores $0.125$ (popularity floor $0.068$; MF cell single-seed). Each class wins its matched task, and only the feature-based tower also wins cold. ``RetrievalFormer loses warm'' is therefore a statement about that protocol's task, not about warm ability in general. The fusion recipe is protocol-specific in the same way: fusing frozen MF onto our model on the next-item benchmark does not help (fused $0.3469$ vs.\ $0.3500$ alone), so the operating point of Table~\ref{tab:coldstart} is a property of matrix-completion-style serving.

\textbf{Within-class ablations (shared embeddings; attention mode).} Within the dual-encoder class, sharing embedding tables across towers matters on average but not at the optimum: marginalized over our MovieLens-1M tuning grid (3 seeds per configuration), separating the towers' embedding tables costs $3.4\%$ Recall@20 and removing shared features entirely costs $6.9\%$; at the single best configuration the gain is within seed noise. Shared embeddings buy robustness (one fewer thing to tune) without driving accuracy; we report marginalized deltas because grid-pooled absolutes depend on grid composition. A second ablation closes the attention-mode axis: replacing prefix-LM attention with pure causal attention (the only change is bidirectionality within the three-token profile/readout prefix, as history attention is causal in both modes) costs $-0.0098$ and $-0.0026$ test Recall@20 on paired seeds (pair mean $0.3627$ vs.\ the constant-LR cap-300 family's $0.3668$), so the bidirectional prefix block is load-bearing at the ${\sim}{+}0.004$ margin and the submitted configuration keeps it.

\end{document}